\documentclass[twocolumn,showkeys,showpacs,aps,nofootinbib,superscriptaddress]{revtex4-1}
\usepackage{color}
\usepackage{bm}% bold math
\usepackage{amssymb}
\usepackage[pdftex]{graphicx}
\usepackage{tabularx}
\usepackage{hyperref}

\usepackage{dcolumn}% Align table columns on decimal point
\usepackage{bm}% bold math
\usepackage{latexsym}
\usepackage{ifthen}
\usepackage{color}

\usepackage{amssymb}

\newcommand{\il}{~}
\begin{document}
\title{A Counterexample  of the MRI in the Two-Dimensional
Axial Symmetry}
\author{Giovanni Montani}
\affiliation{
 ENEA - C.R. Frascati (Rome), UTFUS-MAG.\\
Physics Department, ``Sapienza'' University of Rome,{ c/o VEF, ``Sapienza'' Universit\`a di Roma, P.le Aldo Moro 5, 00185 Roma (Italy)}}
\author{ Daniela Pugliese}
\affiliation{School of Mathematical Sciences, Queen Mary University of London, Mile End Road, London E1 4NS, UK.}
\date{\today}
\begin{abstract}
We analyze a linear perturbation scheme for a two-dimensional
background plasma, which is rotating with a differential
frequency and is embedded in a poloidal magnetic field.
The main  two  assumptions  of the present study, {which  in turn are related,} are that {the plasma profile is axially symmetric, both
in the background and the perturbation approximation,
where  the azimuthal magnetic field is requested to vanish identically
and  secondly }  {the angular frequency on the magnetic surface function only still holds in the
non-stationary regime},
which, on the steady background equilibrium, is ensured
by the validity of the corotation theorem
(Ferraro 1937).
Indeed, such a restriction of the model is rather natural
and it implies that the azimuthal component of the linear
plasma shift is reabsorbed in the expression for the
non stationary electric field (in principle, at any order
of approximation) and can no longer provide a non-zero
azimuthal component of the magnetic tension field.
As a result, the Magneto-rotational instability is
suppressed and the magnetic field has the effect to
stabilize the plasma configuration with respect to the
pure hydrodynamical case.
\end{abstract}
%Astrophysical plasma, 95.30.Qd
%laboratory studies, 52.72.+v
%Hydromagnetic plasma instability, 52.35.Py
%Magnetohydrodynamics
%in astrophysics, 95.30.Qd
%in fluids, 47.35.Tv, 47.65.-d
%in plasma dynamics and flow, 52.30.Cv
%in plasma simulation, 52.65.Kj
%Plasma instabilities, 52.35.-g
%Plasma turbulence, 52.35.Ra
%space plasma, 94.05.Lk
%Plasma waves, 52.35.-g
%Rayleigh-Taylor instabilities, 52.35.Py
\maketitle
\section{Introduction}
The presence of rotation in the quasi-stationary state of
a magnetized plasma is a common feature,
both in astrophysical and
laboratory systems. In stellar accretion disks, the plasma
rotates because of its orbital motion in the gravitational
field of the central body \cite{B01}.
In Tokamak experiments, the
emergence of rotation profiles is observed as spontaneous tendency
of the system evolution and its origin is still questioned,
though three main proposals stand \cite{R07}.

The coupling between the
magnetic field, in which the plasma is embedded, and the differential
rotation of its different layers, lead to the appearance of unstable
modes, named Magneto-rotational instability (MRI),
first discovered by Velikhov in 1959
\cite{V59} and applied to
an astrophysical context by Chandrasekhar in 1960
\cite{C60}.
This effect is relevant in stellar accretion disks, as well as
in Tokamak configurations in order to trigger the onset of a
turbulent behavior in both these scenarios of plasma dynamics.
The merit of having properly characterized
this crucial role of MRI in determining the
transition from a laminar to a disordered
regime of a differentially rotating plasma
is due to Balbus and collaborators in the
early nineteens \cite{BH91} (see also \cite{BH98}).

In the case of a plasma disk,
 surrounding a compact astrophysical object, the presence
of the MRI is the only viable mechanism to generate a turbulent
transport, responsible for accretion, when restated as a laminar
viscoresistive behavior, according to the original idea
for angular momentum
transport \cite{S73}.

In the Tokamak plasma profiles are commonly registered collective
behaviors of the configuration, associated
to both toroidal and poloidal
rotation velocities. Such a phenomenon, dubbed
\emph{spontaneous rotation}, is often favourable to the
plasma confinement, allowing steepen profiles of temperature
and density toward the torus axis.
However the plasma
rotation is naturally suppressed in favor of a turbulent evolution
and, often, it takes place a real interplay between the laminar
and chaotic regimes.
In such a interchange, the MRI is expected to play a role
in originating the unstable modes, then evolving toward
the onset of turbulence, whose associated transport is,
in general, a source of spread for the plasma confinement
\cite{H94}.

For a discussion on the parallelism existing
between the axial symmetry of a stellar accretion
disk and a Tokamak configuration see \cite{C94},
which was recently enforced by the development
of local disk profiles \cite{C05,CR06} characterized
by magnetic field micro-structures resembling
well-known features of the laboratory plasma physics.
For the generalization of such profiles to the global
 radial equilibrium of a steady state of
an accretion disk see \cite{MB011}, while
their impact on the implementation of the Shakura
prescription in a fully magneto-hydrodynamical
scheme is analyzed in \cite{MC012}.

Here, we consider a two-dimensional axisymmetric
configuration of the plasma, built up by a
steady background, characterized by a rotating plasma,
and a first order linear perturbation theory,
which introduces a time dependence in the model.
The normal modes analysis is performed around a
local configuration of the background, determined by
fiducial values of the radial and vertical coordinates.
The analyzed configuration is restricted by requiring
that the angular velocity is, at any order of approximation,
a function of the magnetic surface only.
For the stationary axisymmetric configuration
this feature is assured by the corotation (also known as isorotation) theorem of Ferraro
\cite{F37}, but, in the linear time dependent
perturbation scheme, it plays the role of a specific
assumption on the considered configuration.
Furthermore, we require that the azimuthal component
of the magnetic field identically vanishes. Its exact validity is crucial in the derivation below, however such a
restriction could  be motivated \emph{a posteriori}
both for astrophysical and Tokamak systems.
Starting from these two statements {which, as discussed in Sec.\il(\ref{SeC:aSTRO}), are related } we arrive to
demonstrate that, in such a system, the MRI is
suppressed and that a sufficiently strong magnetic field,
is even able to remove pure hydrodynamical instabilities.

The mechanism responsible for the MRI suppression
is the possibility of reabsorbing the azimuthal component
of the plasma shift into the poloidal electric field.
Indeed, we {are} in practice working in the Coulomb gauge,
dealing with a time-dependent electric potential,
made up of the magnetic surface function only.
Thus, the time dependence of the perturbed magnetic
field arising in the plasma is fixed by the pure
poloidal component of the shift and the azimuthal
component of the magnetic tension vanishes, implying
the MRI suppression.

This result is of impact both in the astrophysical
and laboratory settings, because it shows how,
if the azimuthal magnetic field is sufficiently
small and the angular frequency is forced to
strictly follow the magnetic surface structure,
the fundamental Magneto-hydrodynamical instability,
affecting the plasma rotation profiles,
i.e. the MRI paradigm is naturally removed.
{Indeed in the Tokamak configurations the
transition from spontaneous rotation to
turbulent regimes is expected to take place by
non-linear transitions in regions of the system
parameters (background magnetic field and
perturbation wavelength),
where MRI is anyway suppressed\cite{R07,Terry00,H94,Strait07,Wang11,Kaye12}.
Otherwise the issue of the present analysis strongly concerns
the angular momentum transport in astrophysics
(see the discussion at the end of the paper)}.
In fact, when the MRI is absent in the linear
backreaction of an accreting disk plasma, there exists
no firm alternative scenario to generate the
effective dissipation required by the standard model
for angular momentum balance across the configuration \cite{S73,B01,ShaSun73}.

{
The paper is organized as follows:
In  Sec.\il\ref{Sec:FE}, we give  the fundamental equations of the
axially-symmetric two-dimensional MHD and the assumption setting the accreting  configurations we are considering and in particular  the
relation between the angular frequency of the plasma
and the
magnetic surface  function (isorotation condiction).
In Sec.\il\ref{Sec:Pert}, we develop the perturbation
scheme for the considered problem and we draw  and discuss the main results of this paper.
The phenomenological
implications  and the relevance of these results  in the astrophysical settings are outlined in Sec.\il\ref{SeC:aSTRO}.
Some finally remarks and conclusions  are in Sec.\il\ref{Sec:concl}}

\section{Fundamental equations}\label{Sec:FE}
Let us consider the system
made of the Faraday law and of the electron force balance
\begin{eqnarray}
\nonumber \label{flefb}
\partial _t \vec{B} = - c\vec{\nabla} \wedge \vec{E},
\\
\vec{E} + \frac{\vec{v}}{c}\wedge \vec{B} = 0
\label{flefb2}
\, ,
\end{eqnarray}
where $\vec{E}$ and $\vec{B}$ denote the
electric and the magnetic field respectively.
Equations
(\ref{flefb2}) admit the solution
\begin{equation}
\vec{E} = - \vec{\nabla} \Phi - \frac{1}{c}
\partial _t \vec{A}
\, ,
\label{potvec}
\end{equation}
where $\vec{A}$ is the vector potential,
such that the magnetic field $\vec{B} = \vec{\nabla}\wedge \vec{A}$,
while $\Phi$ denotes the time dependent electric potential
(we are operating in the Coulomb gauge
$\vec{\nabla}\cdot \vec{A} = 0$).
Substituting this solution into the electron
force balance, we get the fundamental relation
\begin{equation}
\vec{\nabla} \Phi + \frac{1}{c}\partial _t \vec{A}
= \frac{\vec{v}}{c}\wedge \vec{B}
\, .
\label{eqf}
\end{equation}
We now specialize this scheme to the two-dimensional
axial symmetry, by assigning a vector potential
of the form
$\vec{A} = (\psi (t, r, z^2)/r) \vec{e} _{\phi}$,
where we adopted cylindrical coordinates
$\{ r, \phi ,z\}$ and we take the magnetic
flux surface function $\psi$ as depending on $z^2$
because the reflection symmetry with respect to
the equatorial plane $z = 0$ is postulated
(for a stellar accretion disk, this is due
to the symmetry of the gravostatic equilibrium).
The {resulting} magnetic field takes the
explicit form
\begin{equation}
\vec{B} = -\frac{1}{r}\partial _z\psi \vec{e}_r +
\frac{1}{r}\partial _r\psi \vec{e}_z
\, .
\label{mgf}
\end{equation}
We also consider the following velocity field
in the plasma
\begin{equation}
\vec{v} = \omega (\psi ) r\vec{e}_{\phi} + \vec{v} _p
\, ,
\label{vlf}
\end{equation}
where $\vec{v} _p = v_r\vec{e}_r + v_z\vec{e}_z$
denotes the poloidal velocity field in the meridian plane.
This choice of the magnetic and velocity fields
contains two main assumptions: the vanishing
nature of the toroidal component $B_{\phi}$ and that
the angular frequency of the disk depends on the
magnetic flux surface, i. e.  $\omega = \omega (\psi )$.
Nonetheless, the considered system is rather general
in axial symmetry
and it does not contain, in principle, any restriction able to suppress
the MRI.
Using the expressions above for $\vec{B}$ and $\vec{v}$
to rewrite the relation (\ref{eqf}), we get
\begin{equation}
\vec{\nabla} \Phi
+ \frac{1}{c}\frac{\partial _t\psi}{r}\vec{e}_{\phi} =
\frac{\omega (\psi )}{c} \vec{\nabla}\psi
{-} \frac{1}{cr}\left( \vec{v}_p\cdot \vec{\nabla} \psi
\right) \vec{e}_{\phi}
\, .
\label{eqf2}
\end{equation}
From the poloidal and azimuthal components
of this equation, we write down
the two key relations
\begin{eqnarray}
\label{eqsf}
\frac{d \Phi}{d \psi} = \frac{\omega (\psi )}{c},
\\
\partial _t\psi = {-}\vec{v}_p\cdot \vec{\nabla} \psi.
\label{eqsf2}
\end{eqnarray}
In what follows, it will be convenient to
use relation
(\ref{eqsf2}) in its
vector form
\begin{equation}
\partial _t \vec{A} = \vec{v}_p\wedge \vec{B}
\, .
\label{seqsf}
\end{equation}
\section{Perturbation scheme}\label{Sec:Pert}
We now consider a linear perturbation scheme to the background
equilibrium of a rotating plasma disk, having angular velocity
$\omega = \omega _0(\psi _0(r_0,z_0^2))$, a mass density
$\rho = \rho_0(r_0,z_0^2)$
and the pressure $p = p_0(r_0,z_0^2)$,
and  embedded in a magnetic field described
by a surface function $\psi _0(r_0,z_0)$
(for a detailed discussion of the static
force balance, see the discussion below,
at the end of the linear perturbation calculus).

Here $r_0$ and $z_0$ are two fiducial values of the radial and vertical coordinates respectively, around which we develop the
local behavior of the perturbations, whose wavelength is
assumed to be sufficiently small to explore only the region
nearby such background circumference.
In this approximation limit, we take the magnetic surfaces in the form
$\psi = \psi _0(r_0,z_0^2) + \psi _1(t,r,z)$, with
$\mid \psi _1\mid \ll \mid \psi _0\mid$ and
\begin{equation}
\mid (\vec{\nabla} \psi _1)\mid
\ll
\mid (\vec{\nabla} _0\psi _0)\mid
\, ,
\label{lincond}
\end{equation}
where $\vec{\nabla}_0$ denotes the gradient operator
with respect to the fixed space coordinates $r_0,z_0$.
Such a condition clearly states that the perturbed magnetic field
is much smaller than the background one, as requested
by the linear approximation.
However, we also require the { natural}
inverse relation for the second order
gradients (current density), i.e.
\begin{equation}
\mid (\vec{\nabla} ^2 \psi _1)
\gg
\mid (\vec{\nabla} ^2_0\psi _0)\mid
\, .
\label{lincond2}
\end{equation}
This paradigm for the gradient fields corresponds
to the natural hierarchy relation in the plasma.
We introduce the poloidal shift vector
$\vec{\xi}_p$
of the plasma,
by defining the corresponding poloidal velocity as
$\vec{v}_p(t,r,z) = \partial _t
\vec{\xi} _p(t,r,z)$.
While, the angular velocity admits the local development
\begin{eqnarray}
\omega = \omega (\psi _0(r_0 - \xi _{pr}, z_0 -\xi_{pz})
+ \psi _1)
\nonumber \\
\simeq \omega _0 - \omega _{0r}^{\prime}\xi_{pr} -
\omega _{0z}^{\prime}\xi_{pz}
+ \dot{\omega}_0 \psi _1
\, ,
\label{omdev}
\end{eqnarray}
where we introduced the notation
\begin{equation}
\omega _{0x}^{\prime} \equiv
\left(\frac{\partial \omega}{\partial x}\right) _{x = x_0},
\dot{\omega}_0 \equiv
\left(\frac{d\omega}{d\psi}\right) _{\psi = \psi _0}.
\label{not}
\end{equation}
From Eq.\il(\ref{omdev}), it is immediate to reach the identification
$\partial _t\xi _{\phi} = \dot{\omega}_0 \psi _1$.
In order to focus our analysis to the main features of the
MRI, we consider incompressible perturbations, i.e. we
require $\vec{\nabla} \cdot \vec{v_1} = 0$ and hence we
get $\vec{\nabla}\cdot \vec{\xi}_p = 0$.
By equation (\ref{seqsf}), we get the following expression for the perturbed
magnetic field
\begin{equation}
\vec{B}_1 = (\vec{B}_0\cdot \vec{\nabla})\vec{\xi}_p
\, .
\label{B1}
\end{equation}
Thus, the magnetic tension force $\vec{T}^{(m)}$
takes the form
(the background pressure and tension are negligible
by means of the inequality (\ref{lincond2}))
\begin{equation}
\vec{T}^{(m)} = \rho _0 v_A^2 (\hat{B}_0\cdot \vec{\nabla})^2\vec{\xi}_p
\, ,
\label{ten}
\end{equation}
here $\hat{B}_0$ is the unit vector in the magnetic
field direction and $v_A\equiv\sqrt{B_0^2/(4\pi \rho_0)}$ denotes the Alfven speed
associated to the background.
Moreover we take the Fourier representation of the perturbation
$\vec{\xi}_p$
\begin{equation}
\vec{\xi}_p =
\vec{\bar{\xi}}_p (t)\exp \{ i(\vec{k}_p\cdot
\vec{r}_p)\}
\, ,
\label{fdec}
\end{equation}
where the index $p$ denotes the poloidal component
of the wave and position vectors. { We note that using  conditions (\ref{lincond}) and (\ref{lincond2})  we have $k_p r_0\gg1$ and $k_p r_0\ll\psi_0/\psi_1$.}
In order to select pure Alfven perturbations,
i.e. to eliminate the magnetic pressure from the problem,
we restrict our attention to wave-vectors of the form
$\vec{k}_p = k\hat{B}_0$, so that, as a consequence of
the incompressibility (namely
$\vec{k}_p\cdot \vec{\xi}_p = 0$), we obtain the
constraint $\hat{B}_0\cdot \vec{\xi}_p = 0$.
Then, the magnetic tension takes the form
\begin{equation}
\vec{T}^{(m)} = -\rho _0k^2v_A^2\vec{\xi}_p
\equiv -\rho_0 \omega _A^2\vec{\xi}_p
\, .
\label{tenfin}
\end{equation}
The momentum conservation equations, for the
velocity perturbations
$\vec{v}_1 = \vec{v}_{p1}
+ v_{\phi 1}\vec{e}_{\phi}$,
take the form
\begin{eqnarray}
\rho _0\left( \partial _tv_{r1} - 2\omega _0 v_{\phi 1}
\right) =
T^{(m)}_r
\label{eu1},\;\\
\rho _0\partial _tv_{z1} = T^{(m)}_z
\label{eu2},\;\\
\rho _0\left( \partial _tv_{\phi 1} + 2\omega _0 v_{r1}
+ r_0\omega ^{\prime}_{0r}v_{r1}
{+r_0 \omega ^{\prime}_{0z}v_{z1}}\right)  = T^{(m)}_{\phi},\;
\label{eu3}
\end{eqnarray}
\cite{B01}. By means of the natural relations
\begin{eqnarray}
v_{r1} = \partial _t\xi _{pr}\, , \quad
v_{z1} = \partial _t\xi _{pz},\\
v_{\phi 1} = {{r_0}}\left(\dot{\omega}_0\psi _1 -
\omega ^{\prime}_{0r}\xi _{pr} -
\omega ^{\prime}_{0z}\xi _{pz}\right),
\,
\end{eqnarray}
and the explicit form
of the magnetic tension,
the momentum conservation equations rewrite as
\begin{eqnarray}
\partial ^2_t\xi _{pr} - 2\omega _0 {r_0}\dot{\omega}_0 \psi _1 =
-(y_r + \omega _A^2) \xi _{pr} -y_z\xi _{pz},
\label{eun1}\\\label{eun2}
\partial ^2_t\xi _{pz} = -\omega _A^2\xi_{pz},
\\
{r_0\dot{\omega}_0\partial _t\psi _1
+ 2\omega _0\partial _t\xi _{pr} = 0}
\label{eun3}
\, ,
\end{eqnarray}
where we adopted the notation
$y_p \equiv 2r_0\omega _0\omega ^{\prime}_{0p}$.
Here, we have retained the full dependence in
$\vec{\xi}_p$ and then the partial time derivative,
since we did not yet Fourier expand $\psi_1$.
The dynamics of the vertical shift $\xi _{pz}$
is fixed by equation (\ref{eun2}) and it yields
a stable behavior of the plasma along such a direction.
Equation (\ref{eun3}) can be easily integrated,
getting the relation
%{
\begin{equation}
r_0\dot{\omega}_0\psi _1 = -2\omega _0\xi _{pr}
\, ,
\label{psi1}
\end{equation}
which substituted into equation (\ref{eun1})
gives the following  equation for
$\bar{\xi} _{pr}$
\begin{equation}
\partial_t^2{\bar{\xi}_{pr}} =
- (\mathcal{K}_0^2  + \omega _A^2) \bar{\xi}_{pr}{-y_z \bar{\xi}_{pz}}
\, ,
\label{rads}
\end{equation}
where $\mathcal{K}_0$ denotes the epicyclic
frequency, i.e.
\begin{equation}
\mathcal{K}_0^2 \equiv
\frac{1}{r_0^3}\partial _{r_0}(\omega _0^2r_0^4) =
4\omega _0^2 + y_r
\, .
\label{epic}
\end{equation}
{Deriving equation\il(\ref{rads}) two times in $t$ and using equation\il(\ref{eun2}) we obtain:
\begin{equation}\label{eq:4-xi-r}
\partial_t^4{\bar{\xi}_{pr}} =
- (\mathcal{K}_0^2  + \omega _A^2) \partial_t^2\bar{\xi}_{pr}{+y_z \omega_A^2\bar{ \xi}_{pz}}
\,.
\end{equation}
Finally using again   equation\il(\ref{rads})  in  equation\il(\ref{eq:4-xi-r})  one has
\begin{equation}\label{eq:4-xi-r+2}
\partial_t^4{\bar{\xi}_{pr}} + (\mathcal{K}_0^2  + 2\omega _A^2) \partial_t^2\bar{\xi}_{pr}{+\omega_A^2 (\mathcal{K}_0^2  + \omega _A^2)\bar{\xi}_{pr}}=0
\,.
\end{equation}}
Taking the following plane-wave representation
for the time dependence of $\bar{\xi}_{pr}$
\begin{equation}
\bar{\xi}_{pr} = \hat{\xi}_{pr}
\exp \{-i\Omega t\},
\label{fom}
\end{equation}
equation {(\ref{eq:4-xi-r+2})} reduces to
{
\begin{equation}
\Omega ^4 -(\mathcal{K}_0^2 +2 \omega _A^2)\Omega^2+\omega_A^2(\mathcal{K}_0^2 + \omega _A^2)=0,
\label{reldisp1}
\end{equation}
admitting  the following solutions
\begin{figure}[h!]
%\centering
\begin{tabular}{c}
\includegraphics[width=1\hsize,clip]{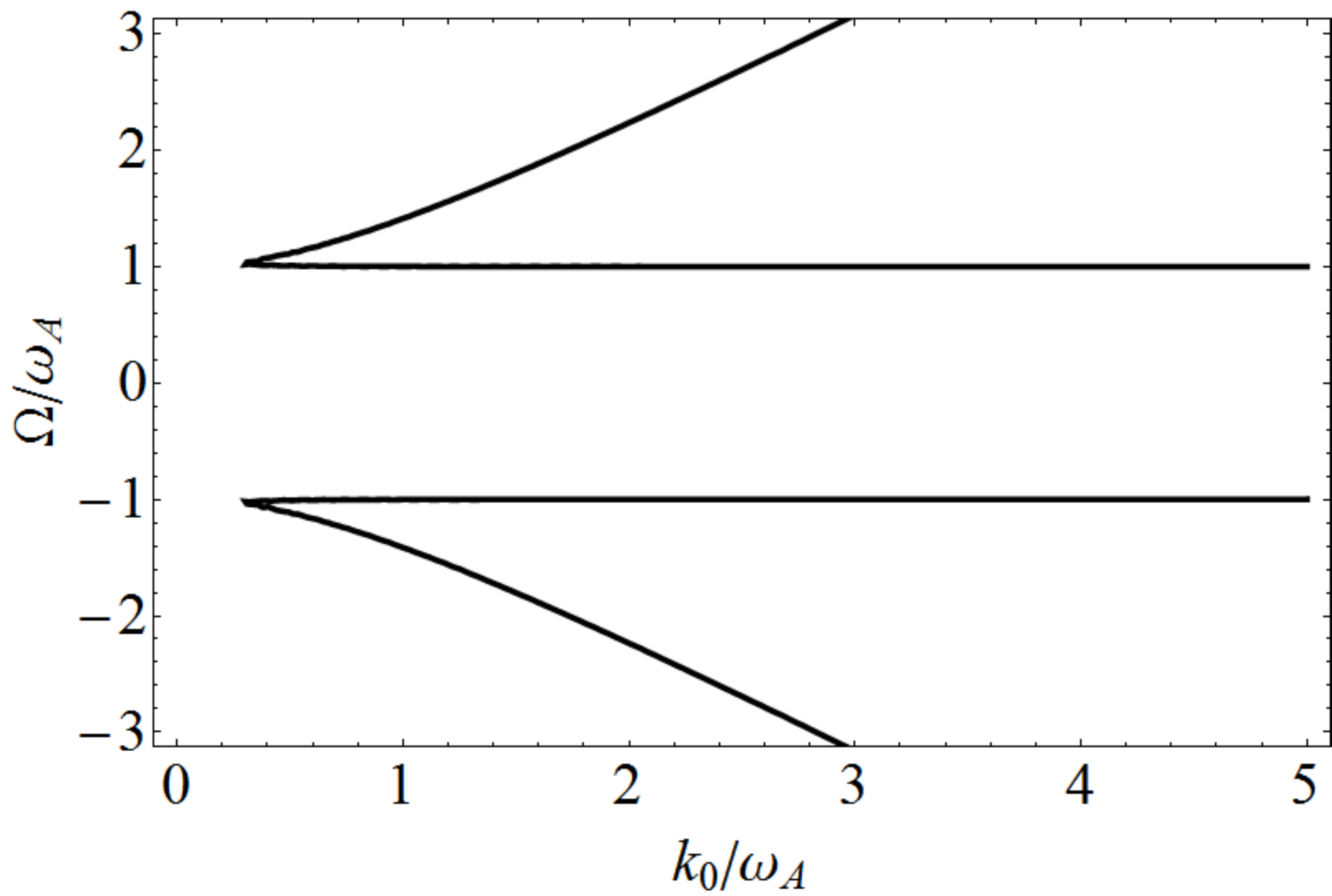}
%(a)&(b)&(c)
\end{tabular}
\caption[font={footnotesize,it}]{\footnotesize{
 Zeros of the fourth degree  polynomial (\ref{reldisp1}) in terms of $\Omega/\omega_A$ and $\mathcal{K}_0/\omega_A$. }}\label{TheZeroSol}
\end{figure}
\begin{equation}\label{reldisp}
\Omega^2=\omega_A^2\quad\mbox{and}\quad\Omega^2=\mathcal{K}_0^2 +\omega _A^2,
\end{equation}}
see also Fig. ({\ref{TheZeroSol}).
In the absence
of the magnetic field, i.e. $\omega_A\equiv 0$,
the dispersion relations (\ref{reldisp}) reduce to
the well-known stability condition
$\mathcal{K}_0^2 > 0$ (see \cite{BH98} for a discussion on the  Rayleigh criterion for local stability).
Such a condition states that the specific
angular momentum of the plasma radially increases and it
holds as far as the angular frequency
sufficiently smooth decays. When the plasma is
magnetized, for having stability,
the relation (\ref{reldisp}) requires
%
%\begin{equation}
$
\mathcal{K}_0^2 + \omega _A^2 > 0
\, .
$
%\label{stabcond}
%\end{equation}
%
Since $\omega _A^2$ is intrinsically positive
by its definition, the effect of the magnetic field
consists of the stabilization of an unstable
part of the spectrum for
the non-magnetized (neutral fluid) configuration, even when the magnetic field is weak.
In this scheme, the MRI is completely suppressed
and replaced by an opposite effect, which stabilizes
the differential rotation profile of the plasma and
prevents the onset of a turbulent behavior.

Let us now analyze if the result above remains valid
in the case of a non-vanishing azimuthal component
of the background magnetic field, i.e.
$B_{0\phi}\neq 0$. Since the (perturbed) poloidal
velocity is a divergenceless field, it can be
expressed via a scalar function
$\Theta(r,z)$ as follows
\begin{equation}
\vec{v}_p = {  \frac{\vec{\nabla}\Theta}{r}}
\wedge \vec{e}_{\phi}
\, .
\label{divlpv}
\end{equation}
Then, in the right-hand side of equation
(\ref{eqf2}) we would have the additional term
$\vec{\nabla}\Theta B_{\phi}/r$ and the possibility
to reabsorb this contribution into the electric
potential, requires that it be a pure gradient.
Thus, we get the following constraint for the azimuthal
magnetic field component
$
B_{\phi} = rf(\Theta )
\, ,
%\label{bcond}
$
$f$ denoting a generic functional form.
To understand if such a form of $B_{\phi}$ is
allowed by the considered plasma configuration,
we have to analyze the background profile.
The magneto-static configuration is fixed by the
basic equation
\begin{equation}
\vec{\nabla}_0p_0 -\rho_0 \omega ^2_0 r_0 \vec{e}_r
= \vec{F}_{0L} - \rho_0\vec{\nabla}_0 \Phi _g
\, ,
\label{mse}
\end{equation}
where
$\vec{F}_L = (\vec{\nabla}\wedge \vec{B})\wedge
\vec{B}/4\pi$
denotes the Lorentz force,
$\Phi _g$ is the gravitational potential
(it concerns the stellar accretion disk configuration only).
By the axial symmetry restriction,
the consistency of the azimuthal component of
equation (\ref{mse}) clearly implies
$F_{0L\phi} = 0$. As it is well-known (and it is
easily checked
by (\ref{mgf})),
this condition implies
$B_{0\phi} = I(\psi _0)/r_0$, here $I$ is a
generic function of the magnetic surface•s.
Assuming this form as valid at any order,
a non-zero azimuthal component of the
magnetic field is allowed in the present scheme,
as far as the condition
$\Theta = g(I/r^2)$ (with $g \equiv f^{-1}$
and $g(I(\psi _0)/r_0^2) = 0$) holds.
In the local model we are considering,
any function $g$ having at least one zero is
suitable for expressing the first order quantity
$\Theta$. It is worth noting that, since $\psi$ is
here taken even in $(z)$, the radial velocity is
odd and vanishes on the plane $z=z_0$, so that
locally we have no real accretion toward the axis
or ejection outward of it.
{{Furthermore}  even if we consider an azimuthal perturbation of the magnetic field as far as its background value is zero, such perturbations are frozen (see eq. (\ref{eqf})).}
Anyway, the possibility to neglect the
azimuthal magnetic field component can also relay
on phenomenological considerations. In Tokamak
configuration this component is much smaller
than the toroidal (here the vertical) one, almost
of the same amount of the ratio between the small
and large radius of the torus, although its  presence  has a  physical role in the plasma  confinement profile.  In the stellar accretion
disks, the azimuthal magnetic field is essentially
due to the dynamo effect. Thus, it is associated to the
turbulent behavior of the plasma, whose
onset is triggered by MRI \cite{PiLeSaLiKoLaz2012,*San-LiLazPiCho2010,*San-LiPiLa2012}. Therefore, in both these
cases we can regard the azimuthal field, if present
with a generic form, as a weak modification,
affecting the picture depicted above.
\section{Astrophysical implementation}\label{SeC:aSTRO}
Despite our results hold for a generic axisymmetric
plasma, its impact is particularly relevant
on the astrophysical setting, in connection with
the accretion process.
Indeed, when, within a binary system, a compact object
acquires material from a less dense massive companion,
the infalling plasma possesses a non-zero angular
momentum and it takes a disk configuration, see for example \cite{juhan-king-Raine2002,Shapiro-2013}.

The present linear stability approach, based
on two-dimensional axial symmetry, applies to thin plasma accretion disks
and it stands also as a valuable scheme for the thick
equilibrium profiles.

In a wide class of systems, especially of stellar
nature (e. g.  X-ray binary stars), the disk configuration is thin, i.e.
the disk depth is mach smaller than the distance from the
centre at which it is estimated, and the two-dimensional
axial symmetry becomes a very good approximation.
It is worth noting that such a symmetry restriction holds
in the case of thick disks too on a morphological level.
Nonetheless, for thick configurations, 3-dimensional
simulation are commonly implemented to determine
the basic phenomena at the ground of the accretion
picture, see for example \cite{Abr.Fra-2013,Fra-Anninos,Igu-Belo-97,Penna:2010hu}.

The role of the MRI in accreting structures is crucial
 as it is commonly accepted as the only reliable mechanism able to generate
a real turbulent behavior. This latter issue
is, in turn, necessary to account for the dissipative
effects requested for the angular momentum transport
toward the accreting source.
The idea that the effective viscosity, associated to
the plasma (MRI triggered) turbulence allows
a non-zero accretion rate of the central object,
was at the ground of the original Shakura model
\cite{S73,ShaSun73} (based on a fluid-like approach) and
{substantiated} by the studies in \cite{BH91,BH98}, which
promoted the MRI to be the main feature   of this scenario.

The present  analysis, clearly states that if the
azimuthal component of the magnetic field
 vanishes, than we
are forced to have $\omega = \omega (\psi )$ and,
hence the MRI is completely suppressed (see below).
This MRI suppression we outlined, has a
very important impact on the accretion morphology
of Neutron Star endowed with a thin accretion disk,
a rather typical situation in X-ray binary systems, see for example \cite{McClintock:2003gx,juhan-king-Raine2002,Abr.Fra-2013} and references.

In fact a very good representation of the magnetic
configuration of a Neutron Star is offered by a dipole
profile, described by the magnetic flux surface
\begin{equation}
\psi = \frac{\mu _0 {r^2}}{\left( r^2 + z^2\right) ^{3/2}}
\quad , \quad \mu _0 = const.
\, ,
\label{dipco}
\end{equation}
where the constant $\mu _0$ defines the magnetic field
amplitude see Figs.\il(\ref{Fig:3D3Dc}).
\begin{figure}[h!]
%\centering
\begin{tabular}{cc}
\includegraphics[width=1\hsize,clip]{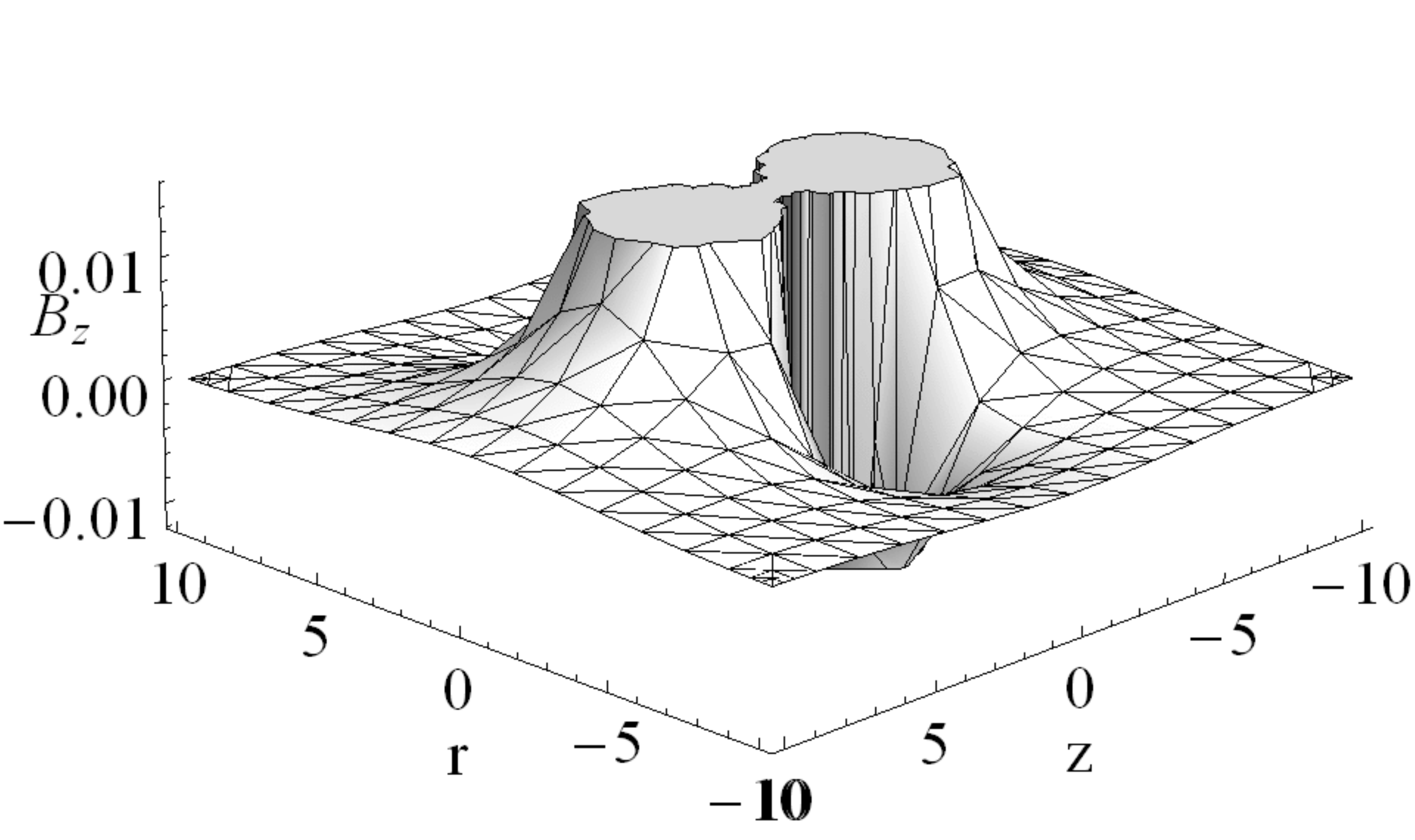}\\
\includegraphics[width=1\hsize,clip]{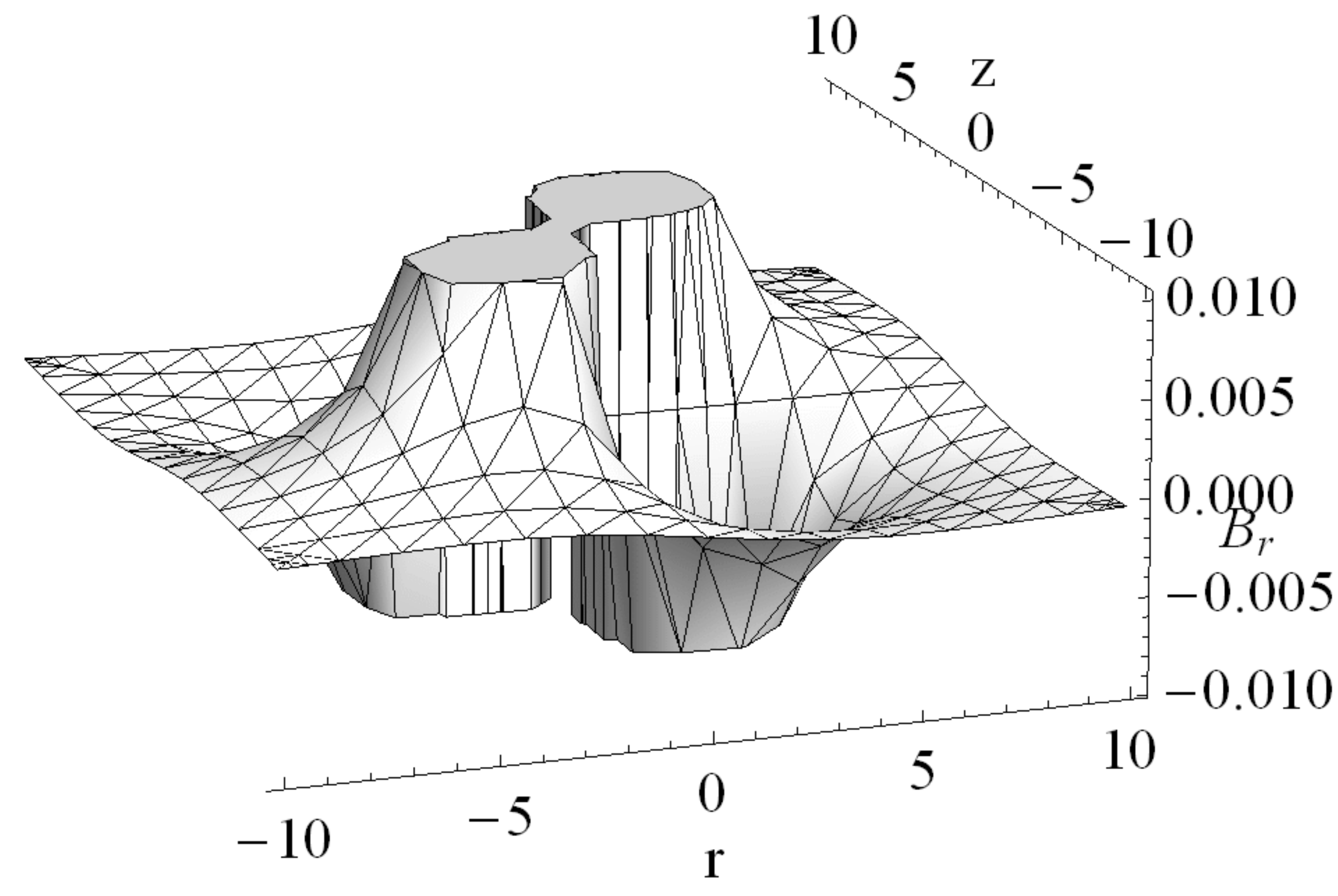}\\
\includegraphics[width=1\hsize,clip]{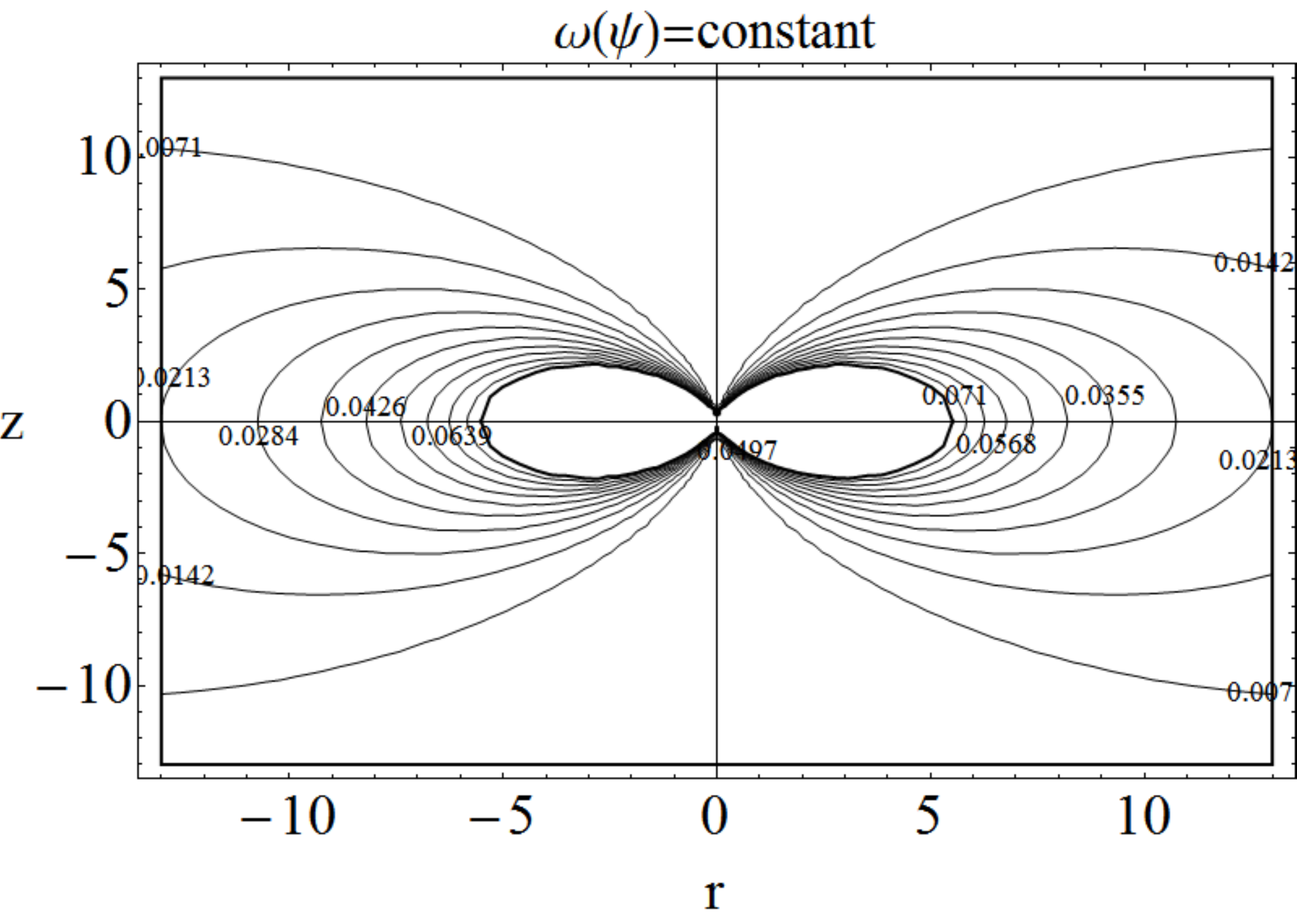}
%\\
%(a)&(b)&(c)
\end{tabular}
\caption[font={footnotesize,it}]{\footnotesize{
Upper and center panels: plots of the poloidal  components of the magnetic field   $B_z$ (left) and $B_r$ (right) as functions of $(r,z)$, for a dipole
profile described by the magnetic flux surface Eq.\il(\ref{dipco}).  Low panel:  the curves  $\omega(\psi)=\mbox{cons}$ as functions of $(r,z)$, where the  angular frequency   is $\omega(\psi)=\sqrt{G \frac{M \psi^3}{{\mu _0}^3}}$, see \cite{MB011}. We have set $G=c=1$.}}\label{Fig:3D3Dc}
\end{figure}

The very striking feature associated to
this configuration is that, in a sufficiently
thin plasma disk, the value of the field can be
properly approximated by its equatorial plane value, i.e.
\begin{equation}
\vec{B}_{eq} \simeq \frac{\mu _0}{r^3}
\vec{e}_z,
\label{dipcoeq}
\end{equation}
and it is easy to check that the azimuthal
component of the induction equation takes
the form
\begin{equation}
\partial _tB_{\phi}  + \partial _r(v_rB_{\phi})
+ \partial _z(v_zB_{\phi}) =
\partial _r\psi \partial _z\omega -
\partial _z\psi \partial _r\omega
\, .
\label{azeq}
\end{equation}

Before discussing the perturbation case,
we observe that, if $B_{\phi} \equiv 0$,
equation (\ref{azeq}) implies
$\omega = \omega (\psi)$, which is a non-stationary
generalization of the corotation Ferraro theorem \cite{F37}.
This fact is relevant because states that the
two assumptions we addressed above are indeed related
and our analysis simply relies on the assumption that
the azimuthal magnetic field vanishes identically.

Viceversa, if $\omega = \omega (\psi )$, the
azimuthal magnetic component satisfies a linear
and homogeneous equation in normal form, which
admits a unique solution. Since the null solution
satisfies the initial condition of a pure poloidal
field, as we considered here, we can easily conclude
that no real mechanism exists, able to generate a magnetic
field in the present scheme.

This fact is of crucial importance because it clarifies that
 {to deal} with a dynamo process, able to generate
a non-zero azimuthal magnetic field, it is
necessary to introduce turbulence effects, like
the so-called $\alpha$-$\omega$  effective terms \cite{StKrRa66,KraRaer80,BraSubra05}.

However, how is it possible to trigger turbulence if
the MRI is suppressed? This scenario seems to
indicate that {there is} a clear difficulty in preserving the
presence of the MRI when the magnetic field of
the central object has a pure poloidal,
in general dipolar, morphology.
This is a rather puzzling scenario for the
astrophysical setting of our study, because
it deprives the standard model for accretion
of the only reliable mechanism, able to trigger the
turbulence behavior of the infalling plasma.
The point is that such a turbulence is necessary
to account the effective dissipation in the plasma disk,
whose presence allows for angular momentum transport
toward the compact object.

In the case of a thin accretion disk, the
quantities $\psi$ and $\omega$ are function of the
radial coordinate only and they are fixed by an average
procedure around the equatorial plane.
In such
a scenario, the right-hand-side of equation (\ref{azeq})
vanishes identically
and therefore it is not immediate to recognize the necessity
 of the isorotation condition.
  In a thin disk, the situation appears
rather simplified (we stress that, in this case, $B_{\phi}=0$ does not implies the isorotation condition) because the magnetic field is
along the vertical direction and our analysis would
reduce to consider the vertical dependence of all
the perturbed fields. Nonetheless the picture
we traced in the general axisymmetric case survives
and it makes very clear the discrepancy between our
analysis and the standard MRI derivation in
\cite{BH91,BH98}.

The origin of the present suppression of the MRI
is in recognizing how when the azimuthal magnetic field
vanishes, the electron force balance
implies the key relation $\omega = \omega (\psi)$,
which is responsible for the different morphology
of the plasma shift vector.

\section{Conclusions}\label{Sec:concl}

In the present analysis, we demonstrate that, in
axial symmetry, conditions exist under which the
MRI is suppressed and it is replaced by a stabilizing
effect of the magnetic field on a differentially
rotating plasma disk.

We start by assuming a vanishing aximuthal magnetic
field and considering an isorotation profile,
although in section\il\ref{SeC:aSTRO} we clarified how for
 { Alfvenic }perturbation, as we considered  here,
the first assumption naturally implies the second one.
In other words, the pure poloidal nature of
the background magnetic field is sufficient to
ensure the validity of the adopted scheme and hence the
removal of the MRI.

The feature from which the MRI suppression originates
{is} the behavior of the azimuthal plasma shift,
which, because of the isorotation condition is
directly related to the correction of the magnetic
flux surfaces, as effect of the plasma backreaction.
In particular, such azimuthal shift enters the electric
field via the electron force balance and therefore it
is removed from the expression of the perturbed
magnetic tension.

After a discussion of the obtained feature in parallel
%\rtb{in both }
between the Tokamak experiments and the astrophysical
setting, we then focus our attention on the latter,
in order to fix the emergence of a puzzling paradigm.
Actually, in the morphology of a stellar accretion disk
features are present, which make up a
crush line in its standard understanding.

The point is that the magnetic field of the central
object has a poloidal behavior, living in the
meridian plane only. Thus, we can naturally consider
the isorotation condition and then we completely
loss the MRI profile of the plasma disk.
However, in this scenario, there is no longer
any mechanism which reliably triggers a full turbulence
behavior, which, in turn, would be able to restore
a non-zero magnetic field.

These considerations open a serious question
about how to restore the MRI, in order to get
the effective visco-resistive effects, which are
required by the standard model of accretion,
accordingly to the original Shakura idea.

The present puzzling theoretical framework calls
attention for a systematic investigation of the
role that each ingredient{s} of the scheme plays
in creating the final picture.
Nonetheless, the present issue suggests that
the standard accretion model is still far from
being a settled down scheme and therefore the
attempts made in \cite{C05,CR06,MB011} in view of a
reformulation of the accretion process have a
rather solid theoretical justification.

\begin{acknowledgments}
This work has been developed in the framework of the CGW Collaboration
(www.cgwcollaboration.it). We would like to thank Nakia Carlevaro for helpful comments and discussions. DP wishes to thank the Blanceflor Boncompagni-Ludovisi, n\'ee Bildt.
\end{acknowledgments}

\end{document}